\documentclass[epj]{svjour}

\usepackage[dvips]{graphicx}

\usepackage{array}
\usepackage{amssymb}
\usepackage{amsmath}
\usepackage{hhline}
\usepackage{longtable}
\usepackage{dcolumn}% Align table columns on decimal point
\usepackage{bm}% bold math
\usepackage{subfigure}
\usepackage{epsfig}
\usepackage{latexsym,amsmath}
\usepackage{amsbsy}

\makeatletter
\renewcommand\appendix{\par
  \setcounter{section}{0}%
  \setcounter{subsection}{0}%
  \setcounter{equation}{0}%
  \renewcommand\theequation{\Alph{section}.\arabic{equation}}
  \renewcommand\thesection{Appendix \@Alph\c@section:}}
\makeatother

\begin{document}
\title{Scaling of mean first-passage time as efficiency measure of nodes sending information on scale-free Koch networks}

\author{Zhongzhi Zhang \thanks{e-mail: zhangzz@fudan.edu.cn} \and Shuyang Gao}                     % Do not remove
\institute{School of Computer Science, Fudan University, Shanghai
200433, China \and Shanghai Key Lab of Intelligent Information
Processing, Fudan University, Shanghai 200433, China}

\date{Received: date / Revised version: date}

\abstract{Random walks on complex networks, especially scale-free
networks, have attracted considerable interest in the past few
years. A lot of previous work showed that the average receiving time
(ART), i.e., the average of mean first-passage time (MFPT) for
random walks to a given hub node (node with maximum degree) averaged
over all starting points in scale-free small-world networks exhibits
a sublinear or linear dependence on network order $N$ (number of
nodes), which indicates that hub nodes are very efficient in
receiving information if one looks upon the random walker as an
information messenger. Thus far, the efficiency of a hub node
sending information on scale-free small-world networks has not been
addressed yet. In this paper, we study random walks on the class of
Koch networks with scale-free behavior and small-world effect. We
derive some basic properties for random walks on the Koch network
family, based on which we calculate analytically the average sending
time (AST) defined as the average of MFPTs from a hub node to all
other nodes, excluding the hub itself. The obtained closed-form
expression displays that in large networks the AST grows with
network order as $N \ln N$, which is larger than the linear scaling
of ART to the hub from other nodes. On the other hand, we also
address the case with the information sender distributed uniformly
among the Koch networks, and derive analytically the global mean
first-passage time, namely, the average of MFPTs between all couples
of nodes, the leading scaling of which is identical to that of AST.
From the obtained results, we present that although hub nodes are
more efficient for receiving information than other nodes, they
display a qualitatively similar speed for sending information as
non-hub nodes. Moreover, we show that that AST from a starting point
(sender) to all possible targets is not sensitively affected by the
sender's location. The present findings are helpful for better
understanding random walks performed on scale-free small-world
networks.
 % end of PACS codes
} %end of abstract

 \maketitle
%%%%%%%%%%%%%%%%%%%%%%%%%%%%%%%%%%%%%%%%%%%%%%%%%%%%%%%%%%%%%%%%%%%%

\section{Introduction}

In recent ten years, as a powerful mathematic tool, as well as a
paradigmatic model in the intense research of complex systems,
complex networks have attracted a surge of interest from the
scientific community~\cite{AlBa02,DoMe02,Ne03,BoLaMoChHw06}. Most
endeavors in the initial few years were devoted to unveil the
nontrivial topological properties of real
systems~\cite{AlBa02,DoMe02}. A lot of empirical studies unraveled
that a large variety of real-life networks display simultaneously
small-world effect~\cite{WaSt98} and scale-free behavior
characterized by a power-law degree distribution~\cite{BaAl99}.
These two important discoveries have radically altered our
understanding for structural aspects of complex networked systems.

After making substantial progress in characterizing the complexity
of real systems, the focus has shifted to dynamical processes
defined on them~\cite{DoGoMe08}, with the aim to uncover the
intrinsic relationship between dynamical processes and underlying
architecture of complex networks, i.e., unravel how deeply the
structural features of networks affect dynamical processes occurring
on them. It has been shown that the power-law degree distribution of
scale-free networks fundamentally influence almost all dynamical
processes taking place on them, such as disease
spreading~\cite{PaVe01}, percolation~\cite{CaNeStWa00},
games~\cite{SaPa05,SaSaPa08}, synchronization~\cite{ArDiKuMoZh08},
to name a few.

In addition to above-mentioned dynamical processes, scale-free
structure also strongly affects the efficiency for random walks with
an immobile trap fixed at a hub node with the highest
degree~\cite{KiCaHaAr08,ZhQiZhXiGu09,ZhGuXiQiZh09,AgBu09,TeBeVo09,AgBuMa10}.
It was surprisingly found that the average receiving time (ART),
i.e., the average of mean first-passage time (MFPT) for a random
walker to a given target hub node, averaged over all source points
in scale-free small-world networks, behaves sublinearly or linearly
with the network order (viz., the number of all nodes). Here the
MFPT from site $u$ to $v$ is defined as the expected time for a
walker starting from $u$ to first reach $v$~\cite{Re01,NoRi04}.
Since the random walker can be looked upon as an information
messenger~\cite{NoRi04,ChBa07}, the low ART to the hub node means
that as information receivers nodes with large degree are efficient
in receiving information. However, any node in a network can also be
treated as an information sender. Then, interesting questions are
raised naturally: What is the scaling of the average sending time
(AST), defined as the average of MFPTs from a hub node to any other
node, chosen uniformly in a scale-free network? Is it still as
efficient as the case that the hub is regarded as a receiver? Does
the location of information sender affect the scaling of AST?
Despite the significance of the questions, they still remain unclear
limited by the difficulty for determining MFPT from a hub node to
some other nodes~\cite{Bobe05}.

In this paper, we study analytically random walks on the class of
Koch networks with scale-free behavior and small-world
effect~\cite{ZhZhXiChLiGu09,ZhGaChZhZhGu10}, which is a fundamental
process gaining considerable recent
attention~\cite{MeKl04,NoRi04a,SoRebe05,BaLo06,CoBeTeVoKl07,GaSoHaMa07,CoTeVoBeKl08,BeChKlMeVo10,BeGrLeLoVo10}.
We first investigate a particular random walk, starting from a hub
node with highest degree to send information to all other nodes,
exclusive the hub itself. We derive exactly the AST from the hub to
another node, averaged over all nodes in the Koch networks. The
obtained explicit formula displays that in large networks with $N$
nodes, the AST grows asymptotically with $N$ as $N \ln N$, which in
sharp contrast to the linear dependence of the ART from all nodes to
the hub~\cite{ZhZhXiChLiGu09}.

In the second part of this work, based on the connection between
random walks and electrical networks, we determine analytically the
global mean first-passage time (GMFPT), defined as the average of
MFPTs over all node pairs. We present that the GMFPT is also
asymptotic to $N \ln N$. Since the GMFPT can be looked upon as the
average of ASTs with the sender distributed uniformly among all
nodes, we conclude that neither the structure inhomogeneity nor the
position of starting points has an essential effect on the scaling
of AST in Koch networks. Thus, the $N \ln N$ behavior of AST from a
particular sender is a representative property of the Koch networks,
which is in comparison with the trapping problem, where the scaling
of ART to a trap (information receiver) depends on the location of
the trap~\cite{TeBeVo09}.

%%%%%%%%%%%%%%%%%%%%%%%%%%%%%%%%%%%%%%%%%%%%%%%%%%%%%%%%%%
% Figure  1
%%%%%%%%%%%%%%%%%%%%%%%%%%%%%%%%%%%%%%%%%%%%%%%%%%%%%%%%%%
\begin{figure}
\begin{center}
\includegraphics[width=0.65\linewidth,trim=100 10 100 0]{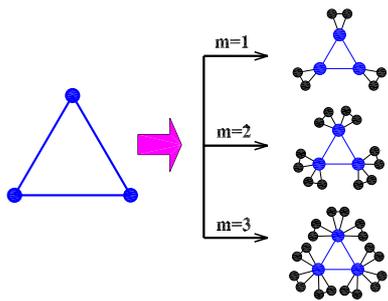}
\end{center}
\caption[kurzform]{\label{iterative} (Color online) Iterative
construction method for the Koch networks. }
\end{figure}
%%%%%%%%%%%%%%%%%%%%%%%%%%%%%%%%%%%%%%%%%%%%%%%%%%%%%%%%%%

\section{\label{sec:net}Koch networks and their structural properties}

\begin{figure}
\begin{center}
\includegraphics[width=0.9\linewidth,trim=100 10 100 0]{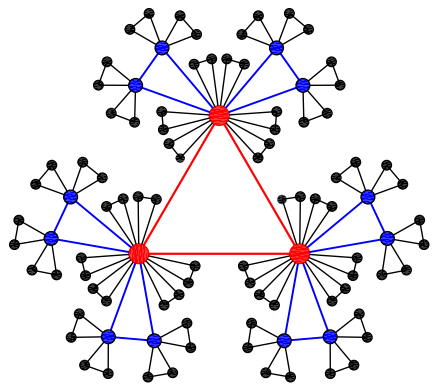}
\end{center}
\caption[kurzform]{\label{network2}(Color online) A network
corresponding to the case of $m=2$.}
\end{figure}

The family of Koch networks controlled by a positive integer
parameter $m$ are translated from the famous Koch
fractals~\cite{LaVaMeVa87} and can be built in an iterative
way~\cite{ZhZhXiChLiGu09,ZhGaChZhZhGu10}. Denote by $K_{m,t}$ the
Koch network family after $t$ iterations. Then, the Koch networks
can be created in the following way: Initially ($t=0$), $K_{m,0}$
consists of three nodes forming a triangle. For $t\geq 1$, $K_{m,t}$
is obtained from $K_{m,t-1}$ by adding $m$ groups of nodes for each
of the three nodes of every existing triangle in $K_{m,t-1}$. Each
node group includes two nodes, both of which and their ``mother''
node are linked to each other constituting a new triangle. In other
words, in order to get $K_{m,t}$ from $K_{m,t-1}$, one can
substitute a connected cluster on the right-hand side (rhs) of arrow
in Fig.~\ref{iterative} for each triangle in $K_{m,t-1}$.
Figure~\ref{network2} illustrates a Koch network for the case of
$m=2$ after several iterations.

%%%%%%%%%%%%%%%%%%%%%%%%%%%%%%%%%%%%%%%%%%%%%%%%%%%%%%%%%
% Figure  2
%%%%%%%%%%%%%%%%%%%%%%%%%%%%%%%%%%%%%%%%%%%%%%%%%%%%%%%%%%
%%%%%%%%%%%%%%%%%%%%%%%%%%%%%%%%%%%%%%%%%%%%%%%%%%%%%%%%%%

By construction, we can obtain with ease some quantities that will
be very useful for deriving the basic quantity we are concerned in
this paper. It is obvious that the number of triangles
$L_\triangle(t)$ present at iteration $t$ is
$L_\triangle(t)=(3m+1)^t$, and the number of nodes generated at
iteration $t$ is $L_v(t)=6m\,L_\triangle(t-1)=6m\,(3m+1)^{t-1}$.
Then, the numbers of edges and nodes in networks $K_{m,t}$ are
\begin{equation}\label{Et}
E_t=3\,L_\triangle(t)=3(3m+1)^t\,
\end{equation}
and
\begin{equation}\label{Nt}
N_t=\sum_{t_i=0}^{t}L_v(t_i)=2\,(3m+1)^{t}+1\,,
\end{equation}
respectively.

Denote by $k_i(t)$ the degree of a node $i$ at iteration $t$ that
entered the networks at iteration (step) $t_i$ ($t_i\geq 0$). Then,
$k_i(t_i)=2$. Denote by $L_\triangle(i,t)$ the number of triangles
passing by node $i$ at step $t$. According to the network generation
algorithm, each triangle passing node $i$ at a given step will lead
to $m$ new triangles involving $i$ at next time step. Hence,
$L_\triangle(i,t)=(m+1)\,L_\triangle(i,t-1)=(m+1)^{t-t_{i}}$. In
addition, the relation $k_i(t)=2\,L_\triangle(i,t)$ holds. Then $
k_i(t)=2(m+1)^{t-t_{i}}$ that indicates
\begin{equation}\label{ki2}
k_i(t)=(m+1)\,k_i(t-1).
\end{equation}
Note that in $K_{m,t}$ the initial three nodes created at iteration
0 have the highest degree $2(m+1)^{t}$. We call these nodes hub
nodes and label by 1 one of the hub nodes, while label the other two
hubs by 2 and 3, respectively.

The Koch networks exhibit some classic characteristics of real-life
systems~\cite{ZhZhXiChLiGu09,ZhGaChZhZhGu10}. They are scale-free
with their degree distribution $P(k)$ following a power-law form
$P(k)\sim k^{-\gamma}$, where $\gamma$ is equal to
$1+\frac{\ln(3m+1)}{\ln(m+1)}$ belonging to the interval $[2,3]$.
They display small-world effect with a small average path length
(APL) and a large clustering coefficient. Their APL exhibits a
logarithmic scaling with network order $N_t$.

%The exact solution to APL in $K_{m,t}$ is
%\begin{equation}\label{APL}
%d_t = \frac {3m+5+(24mt+24m+4)(3m+1)^{t}}{3(3m+1)\,[2(3m+1)^t+1]},\,
%\end{equation}
%which approximates $\frac{4mt}{3m+1}$ in the infinite $t$ limit. As
%shown in Eq.~(\ref{Nt}), the network order $N_t$ grows exponentially
%with $t$, the APL $d_t$ thus shows a logarithmic scaling with $N_t$.

\section{\label{sec:standwalk}Basic properties of Random walks on Koch networks}

After introducing the Koch networks $K_{m,t}$ and their topological
features, we proceed to study standard random walks~\cite{NoRi04}
running on $K_{m,t}$. At each step the walker, located on a given
node, moves uniformly to any of its nearest neighbors. Our main aim
is to find the AST from one of the three hub nodes (e.g., node 1) to
another node averaged over all target nodes except the hub node
itself. To achieve this goal, we provide some essential properties
for random walks on the Koch networks.

\subsection{Evolutionary rule for mean first-passage time}

We fist establish the scaling relation governing the evolution of
MFPT between an arbitrary pair of two nodes, using the approach
based on underlying backward
equations~\cite{Bobe05,HaBe87,ZhZhZhYiGu09}. Let $F_{ij}(t)$ express
the MFPT of the walker in networks $K_{m,t}$, starting from node $i$
to visit node $j$ for the first time. Because of the particular
construction the Koch networks, the exact relation governing
$F_{ij}(t+1)$ and $F_{ij}(t)$ can be given.

Consider an arbitrary node $i$ in the Koch networks $K_{m,t}$ after
$t$ iterations. Equation~(\ref{ki2}) indicates that upon growth of
the networks from generation $t$ to $t+1$, the degree $k_i(t)$ of
node $i$ grows by $m$ times, i.e., it increases from $k_i(t)$ to
$(m+1)k_i(t)$. Denote by $X$ the MFPT from node $i$ to any of its
$k_i$ old neighbors belonging to $K_{m,t}$, and denote by $Y$ MFPT
for a walker starting from any of the $mk_i$ new neighbors of $i$
created at iteration $t+1$ to one of its $k_i$ old neighboring nodes
previously existing before iteration $t+1$. Then the following
simultaneous equations hold:
\begin{eqnarray}\label{FPT1}
\left\{
\begin{array}{ccc}
X&=&\frac{1}{m+1}+\frac{m}{m+1}(1+Y),\\
Y&=&\frac{1}{2}(1+X)+\frac{1}{2}(1+Y),
 \end{array}
 \right.
\end{eqnarray}
which result in $X=3m+1$. Thus, when the networks grow from
generation $t$ to $t+1$, the MFPT from any node $i$ ($i \in
K_{m,t}$) to any node $j$ ($j \in K_{m,t+1}$) increases on average
$3m$ times, namely,
\begin{equation}\label{FPT2}
F_{ij}(t+1)=(3m+1)\,F_{ij}(t)\,.
\end{equation}
This scaling is a basic property of random walks on the Koch
networks, which is very useful for deriving our main result.

\subsection{Scaling relation and expression for average return time}

Let $R_{i}(t)$ denote the expected time for a walker in networks
$K_{m,t}$ originating from node $i$ to return to the starting point
$i$ for the first time, named mean return time (MRT) in the
following text. By definition, we have
\begin{equation}\label{FPT3}
R_{i}(t)=\frac{1}{k_i(t)}\sum_{j \in \Omega_i^{(t)}}[1+F_{ji}(t)]\,,
\end{equation}
where $\Omega_i^{(t)}$ is the set of neighbors of node $i$, which
belong to $K_{m,t}$.

On the other hand, for $K_{m,t+1}$,
\begin{equation}\label{FPT4}
R_{i}(t+1)=\frac{m}{m+1}\times
3+\frac{1}{m+1}\frac{1}{k_i(t)}\sum_{j \in
\Omega_i^{(t)}}[1+F_{ji}(t+1)]\,,
\end{equation}
which can be elaborated as follows. The first term on the rhs of
Eq.~(\ref{FPT4}) describes the process where the walker moves from
node $i$ to its new neighbors and back. Since among all $i$'s
neighbors belonging to $K_{m,t+1}$, $\frac{m}{m+1}$ of them are new,
such a process happens with a probability of $\frac{m}{m+1}$ and
takes three time steps. The second term on the rhs of
Eq.~(\ref{FPT4}) accounts for the process in which the walker steps
from $i$ to one of the old neighbors $j$ previously existing in
$K_{m,t}$ and back; this process occurs with the complimentary
probability $\frac{1}{m+1}=1-\frac{m}{m+1}$. Using Eqs.~(\ref{FPT2})
and (\ref{FPT3}) to simplify Eq.~(\ref{FPT4}), we can obtain the
following relation
\begin{equation}\label{FPT5}
R_{i}(t+1)= \frac{3m + 1}{m + 1}R_{i}(t)\,.
\end{equation}

We next determine the MRT for an arbitrary newly born node in
$K_{m,t}$ that is generated at iteration $t$. Let $i'$ be a new
neighbor of an old node $i$ existing in $K_{m,t-1}$, which is
created at iteration $t$. Note that when $i'$ was generated, another
new node
$i''$ appeared at the same time and is linked to $i$ and $i'$. %Let $A$ denote the FPT from $i$ to $i'$, and
Let $A$ express the MRT of a walker starting off from $i$ in
networks $K_{m,t}$ without ever visiting $i'$ and $i''$. Then we
have the following relations
\begin{equation}\label{RFPT01}
R_{i'}(t) = \frac{1}{2}[1+F_{ii'}(t)] + \frac{1}{2}[1+
F_{i''i'}(t)]\,,
\end{equation}
\begin{equation}\label{RFPT02}
F_{i''i'}(t) = \frac{1}{2} \times 1 + \frac{1}{2}[1+F_{ii'}(t)]\,,
\end{equation}
and
\begin{equation}\label{RFPT03}
F_{ii'}(t) = \frac{1}{k_i(t)} + \frac{1}{k_i(t)}[1 + F_{i''i'}(t)] +
\frac{k_i(t) - 2}{k_i(t)}[A+F_{ii'}(t)].
\end{equation}

The three terms on the rhs of Eq.~(\ref{RFPT03}) can be understood
based on the following three processes: with probability
$\frac{1}{k_i(t)}$, the walker gets from node $i$ to $i'$ in one
time step; with probability $\frac{1}{k_i(t)}$, the walker reaches
node $i''$ in one time step then takes time $F_{i''i'}(t)$ to visit
$i'$; and with the remaining probability $\frac{k_i(t)-2}{k_i(t)}$,
the walker selects uniformly a neighbor node except $i'$ and $i''$
and spends on average time $A$ in returning to $i$ then takes time
$F_{ii'}(t)$ to arrive at node $i'$.

In order to close Eqs.~(\ref{RFPT01}) and (\ref{RFPT03}), we write
the MRT of node $i$ as:
\begin{equation}\label{RFPT04}
R_{i}(t) = \frac{1}{k_i(t)} \times 3 + \frac{1}{k_i(t)} \times 3 +
\frac{k_i(t)-2}{k_i(t)}\times A\,.
\end{equation}
The first (second) on the rhs of Eq.~(\ref{RFPT04}) describes the
process that the walker steps from $i$ to $i'$ ($i''$) and back,
which occurs with probability $\frac{1}{k_i(t)}$ and needs three
time steps. The explanation of the third term is analogous to that
of Eq.~(\ref{RFPT03}).

Eliminating the three intermediate quantities $F_{ii'}(t)$,
$F_{i''i'}(t)$, and $A$, we have
\begin{equation}\label{RFPT05}
R_{i'}(t)=\frac{k_i(t)}{2}\,R_{i}(t).
\end{equation}
Combining Eqs.~(\ref{FPT5}) and (\ref{RFPT05}) and considering
$k_i(t)=2(m+1)^{t-t_i}$ lead to the following closed-form expression
\begin{equation}\label{RFPT06}
R_{i'}(t)=3(3m+1)^t.
\end{equation}
Note that Eq.~(\ref{RFPT06}) can also be obtained from the Kac
formula~\cite{AlFi99,SaDoMe08}, which states that the MRT for a node
is in fact the inverse probability to find a particle at this node
in the final equilibrium state of the random-walk process.

Equation~(\ref{RFPT06}) does not depend the degrees of the old
nodes, to which the new nodes $i'$ is connected, which means that
all the simultaneously emerging new nodes have identical MRT. Since
all nodes born at the same time step have identical degree, this is
obvious from the Kac formula: for any node with degree $k$ in
$K_{m,t}$, its MRT is $\frac{2E_t}{k}$, which is consistent with
Eq.~(\ref{RFPT06}) and in turn implies that Eq.~(\ref{RFPT06}) is
right.

\section{\label{sec:standwalk2} Average sending time from a hub node to another node selected uniformly in the Koch networks}

In this section, we investigate the AST from a hub node to another
node distributed uniformly in the Koch networks. We focus on the
case that the starting point is the hub node 1. Notice that, due to
the symmetry, the starting position can be also node 2 or node 3,
which does not have any influence on the AST. In what follows, we
will show that the particular selection of the starting point makes
it possible to derive analytically the relevant quantity, i.e., AST
from node 1 to all other nodes. Let $T_i(t)$ express the MFPT of
node $i$ in $K_{m,t}$, which is the expected time for a walker
starting from node $1$ to first hit node $i$. The average of MFPT
$T_i(t)$ over all target nodes in $K_{m,t}$ is AST, presented by
$\langle T \rangle_t$, the explicit determination of whose solution
is a main goal of the following text.

For the sake of convenient description for calculating $\langle T
\rangle_t$, we use $\Delta_t$ to denote the set of nodes in
$K_{m,t}$, and use $\bar{\Delta}_t$ to present the set of nodes
created at generation $t$. Thus, we have $\Delta_t=\bar{\Delta}_t
\cup \Delta_{t-1}$. By definition, the quantity concerned $\langle T
\rangle_t$ can be defined as
\begin{equation}\label{MFPT01}
\left \langle T \right \rangle _t =
\frac{1}{{N_t}-1}T_{\text{tot}}(t)\,,
\end{equation}
where $T_{\text{tot}}(t)$ is the sum of MFPTs for all nodes starting
from the hub node 1, i.e.,
\begin{equation}\label{MFPT02}
T_{\text{tot}}(t)=\sum_{i \in \Delta_t} T_i(t)\,.
\end{equation}

Thus, the problem of determining $\langle T \rangle_t$ is reduced to
finding $T_{\text{tot}}(t)$. Since all nodes in $K_{m,t}$ belong to
either $\Delta_{t-1}$ or $\bar{\Delta}_t$, $T_{\text{tot}}(t)$ can
be written as the sum of the two following terms:
\begin{equation}\label{TMFPT03}
T_{\text{tot}}(t) = \sum_{j' \in \bar{\Delta}_t} T_{j'}(t) + \sum_{j
\in \Delta_{t - 1}} {{T_{j}}(t)}\,.
\end{equation}
Using the relation provided by Eq.~(\ref{FPT2}), Eq.~(\ref{TMFPT03})
can be recast as
\begin{equation}\label{TMFPT04}
T_{\text{tot}}(t)  = \sum_{j' \in \bar{\Delta}_t} T_{j'}(t)+
(3m+1)T_{\text{tot}}(t-1)\,.
\end{equation}

Hence, to calculate $T_{\text{tot}}(t)$, one only need to evaluate
the first term on the rhs of Eq.~(\ref{TMFPT04}), which accounts for
the sum of the MFPTs from node 1 to all newly generated nodes at
step $t$. Since before visiting node $j'$ for a walker starting from
node $1$, it must first arrive at node $j$ (an old neighbor of $j'$)
that previously existed at step $t-1$, then $T_{j'}(t)$ can be
written as:
\begin{equation}\label{TMFPT05}
T_{j'}(t) = T_{j}(t) + F_{jj'}(t)\,.
\end{equation}

Next we will show that $F_{jj'}(t)$ can be expressed in terms of the
quantity $R_{j'}(t)$ that has been determined in preceding section.
Note that when $j'$ was born, it was linked to node $j$ and a
simultaneously emerging node $j''$ that was also connected to $j$,
then we have the following useful relations:
\begin{equation}\label{TMFPT06}
R_{j'}(t) =\frac{1}{2}[1+F_{jj'}(t)] + \frac{1}{2}[1 +
F_{j''j'}(t)]\,,
\end{equation}
and
\begin{equation}\label{TMFPT07}
F_{j''j'}(t) = \frac{1}{2} + \frac{1}{2}[1 +F_{jj'}(t)]\,.
\end{equation}
Plugging Eq.~(\ref{TMFPT07}) into Eq.~(\ref{TMFPT06}) leads to
\begin{equation}\label{TMFPT08}
F_{jj'}(t) = \frac{4}{3}R_{j'}(t) - 2\,.
\end{equation}
Inserting the obtained result for $F_{jj'}(t)$ given in
Eq.~(\ref{TMFPT08}) into Eq.~(\ref{TMFPT05}), we obtain
\begin{equation}\label{TMFPT09}
T_{j'}(t) = T_{j}(t)+\frac{4}{3}R_{j'}(t)- 2\,.
\end{equation}

With the result given by Eq.~(\ref{TMFPT09}), the first term on the
rhs of Eq.~(\ref{TMFPT04}), denoted by ${T}_{\text{tot}}^{(1)}(t)$,
can be expressed as
\begin{equation}\label{TMFPT10}
{T}_{\text{tot}}^{(1)}(t)=\sum_{j' \in \bar \Delta_t} T_{j'}(t) =
\sum_{j' \in \bar \Delta_t} \left(T_{j}(t)+
\frac{4}{3}R_{j'}(t)-2\right).
\end{equation}
Since for any node $j$ created at step $t_j$ that belongs to
$\Delta_{t-1}$, there are $L_\triangle(j,t-1)=(m+1)^{t-t_j-1}$
triangles passing by $j$, each of which will lead to $2m$ new nodes
connecting $j$ at step $t$, then using Eqs.~(\ref{RFPT06}) and
Eq.~(\ref{TMFPT10}), the sum $T_{\text{tot}}^{(1)}(t)$ can be
rewritten as
\begin{eqnarray}\label{TMFPT11}
T^{(1)}_{\text{tot}}(t)&=& \sum_{j \in \Delta_{t-1}} 2m L_\triangle(j,t-1) T_j(t)+\nonumber\\
&\quad& ({N_t} - {N_{t - 1}})\left[\frac{4}{3} \times 3{(3m + 1)^t}
- 2\right]\nonumber\\
&=& \sum_{j \in \Delta_{t-1}} 2m (m+1)^{t-t_j-1} T_j(t)+\nonumber\\
&\quad& ({N_t} - {N_{t - 1}})\left[\frac{4}{3} \times 3{(3m + 1)^t}
- 2\right]\,.
\end{eqnarray}

The second term on the rhs of Eq.~(\ref{TMFPT11}) is easy to
compute. So, we only need to work out the first term on the rhs of
Eq.~(\ref{TMFPT11}), represented by $T_{\text{sum}}(t)$, namely,
$T_{\text{sum}}(t)=\sum_{j \in \Delta_{t-1}} 2m (m+1)^{t-t_j-1}
T_j(t)$. Evidently, we have the following recursive relation
\begin{eqnarray}\label{TMFPT12}
&\quad&T_{\text{sum}}(t) \nonumber\\
&=& (3m+1)(m+1)\sum_{j \in \Delta_{t-2}} 2m(m+1)^{t-t_j-2} T_j(t-1) \nonumber\\
&\quad& + 2m(3m+1)\sum_{j \in \bar{\Delta}_{t-1}} T_{j}(t-1)\nonumber\\
&=&
(3m+1)(m+1)T_{\text{sum}}(t-1)+2m(3m+1)T^{(1)}_{\text{tot}}(t-1)\,. \nonumber\\
\end{eqnarray}

On the other hand, Eq.~(\ref{TMFPT11}) can be rewritten as
\begin{equation}\label{TMFPT15}
T^{(1)}_{\text{tot}}(t)= T_{\text{sum}}(t)+ ({N_t} - {N_{t -
1}})\left(4{(3m + 1)^t} - 2\right)\,.
\end{equation}
Considering the initial conditions
$T^{(1)}_{\text{tot}}(1)=96m^2+20m$ and
$T_{\text{sum}}(1)=24m^2+8m$, we can solve recursively the
simultaneous equations~(\ref{TMFPT12}) and (\ref{TMFPT15}) to obtain
\begin{equation}\label{TMFPT17}
T_{\text{sum}}(t)
=8m\left[(3m+1)^{t-1}+3m(2t-1)(3m+1)^{2(t-1)}\right],
\end{equation}
and
\begin{eqnarray}\label{TMFPT18}
T^{(1)}_{\text{tot}}(t)&=&4m(3m+1)^{t-2}\big[6(2mt+2m+1)(3m+1)^{t}\nonumber\\
&\quad&-3m-1 \big].
\end{eqnarray}

Inserting Eq.~(\ref{TMFPT18}) into Eq.~(\ref{TMFPT04}), we can solve
Eq.~(\ref{TMFPT04}) inductively to yield
\begin{eqnarray}\label{TMFPT19}
T_{\text{tot}}(t) &=& \frac{4}{3}{(3m + 1)^{t - 1}}[(12mt + 12m +
2){(3m + 1)^t}\nonumber\\
 &\quad&- 3mt - 3m + 1].
\end{eqnarray}
Inserting Eq.~(\ref{TMFPT19}) into Eq.~(\ref{MFPT01}), we obtain the
explicit expression for the AST $\left\langle T \right\rangle_t$:
\begin{equation}\label{TMFPT20}
\left\langle T \right\rangle_t = \frac{2}{{3(3m + 1)}}[(12mt + 12m +
2){(3m + 1)^t} - 3mt - 3m + 1]\,.
\end{equation}

We continue to show how to express the key quantity $\langle T
\rangle_t$ in terms of the network order $N_t$, in order to obtain
the relation between these two quantities. Recalling Eq.~(\ref{Nt}),
we have $(3m+1)^t=(N_t-1)/2$ and $t=[\ln(N_t-1)-\ln2]/\ln(3m+1)$.
Thus, Eq.~(\ref{TMFPT20}) can be further expressed as a function of
$N_t$ as
\begin{eqnarray}\label{TMFPT21}
\langle T \rangle_t&=&\frac{{{N_t} - 1}}{{3(3m + 1)}}\left(\frac{{12m[\ln ({N_t} - 1) - \ln 2]}}{{\ln (3m + 1)}} + 12m + 2\right) - \nonumber\\
&\quad&\frac{2}{{3(3m + 1)}}\left(\frac{{3m[\ln ({N_t} - 1) - \ln 2]}}{{\ln (3m + 1)}} + 3m - 1\right).\nonumber\\
\end{eqnarray}
Thus, for large networks,
\begin{equation}\label{TMFPT22}
{\left\langle T \right\rangle _t} \sim
\frac{4m}{(3m+1)\ln(3m+1)}{(N_t-1)}\ln {(N_t-1)}\,,
\end{equation}
showing that the AST grows with increasing order $N_t$ as $N_t \ln
N_t$. This leading asymptotic $N_t \ln N_t$ dependence of AST on the
network order is in contrast with the linear scaling of receiving
efficiency on network order for a receiver located at the same hub
node receiving information sent from all other different
nodes~\cite{ZhZhXiChLiGu09}.

It is known that the exponent $\gamma$ of degree distribution for a
scale-free network characterizes the inhomogeneity of the network,
which often strongly affects the dynamical processes running on the
network~\cite{Ne03,BoLaMoChHw06,DoGoMe08}. As shown in
section~\ref{sec:net}, the exponent in the Koch networks is
$\gamma=1+\frac{\ln(3m+1)}{\ln(m+1)}$, implying that parameter $m$
controls the extent of heterogeneous structure of the Koch networks:
the larger the value of $m$, the more heterogeneous the networks.
However, as shown in Eq.~(\ref{TMFPT22}), although for different $m$
the AST of whole family of Koch networks is quantitatively
different, it exhibits the same scaling behavior despite the
distinct extent of structure inhomogeneity of the networks
corresponding to $m$. %, meaning that the structure heterogeneity of the networks has no essential impact on the scaling of the PMFPT.

\section{\label{sec:standwalkB} Global mean first-passage time for the broadcaster uniformly distributed among all nodes}

In the previous section, we have presented that the AST from a most
connected node to another node, averaged over all possible target
points, exhibits a linear dependence with network order by a
logarithmic correction. However, for this case, the information
sender is placed on a largest node. Then a question arises naturally
whether this scaling is representative. Another interesting issue is
whether the diffusion speed still follows the same behavior when the
sender is located on other nodes. In the following text, we will
study the case that the information sender is uniformly distributed
among all nodes in the networks, in order to explore how deeply the
position of the sender affect the scaling of transportation
efficiency.

\subsection{\label{sec:standwalkB01} Exact solution to global mean first-passage time}

In this case, we are concerned in a new quantity called global mean
first-passage time (GMFPT), which is the average of mean
first-passage times over all pairs of nodes in the networks.
Concretely, the GMFPT in $K_{m,t}$, represented by $\langle F
\rangle_t$, is defined as
\begin{equation}\label{GMFPT01}
\langle F \rangle_t=\frac{F_{\rm
tot}(t)}{N_t(N_t-1)}=\frac{1}{N_t(N_t-1)}\sum_{j=
1}^{N_t}\sum_{\stackrel{i=1}{ i \ne j}}^{N_t}F_{ij}(t)\,,
\end{equation}
in which
\begin{equation}\label{GMFPT02}
F_{\rm tot}(t)=\sum_{j= 1}^{N_t}\sum_{\stackrel{i=1}{ i \ne
j}}^{N_t}F_{ij}(t)
\end{equation}
is the sum of MFPTs between all pairs of nodes. Note that the
definition of GMFPT involves a double average: The first one is over
all the walkers to a given target (receiver) $j$, the second one is
over a uniform distribution of target nodes among all nodes in
$K_{m,t}$.

It should be noticed that the above method used for computing
$\langle T \rangle_t$ is not applicable to $\langle F \rangle_t$, so
we must resort to an alternative approach. Fortunately, the peculiar
construction of the Koch networks and the
link~\cite{ChRaRuSm89,Te91} between effective resistance and the
MFPTs for random walks allow to calculate analytically GMFPT
$\langle T \rangle_t$. We view $K_{m,t}$ as resistor
networks~\cite{DoSn84} by considering each edge to be a unit
resistor. Let $R_{ij}(t)$ be the effective resistance between two
nodes $i$ and $j$ in the electrical networks obtained from
$K_{m,t}$. Then, according to the relation between MFPTs and
effective resistance~\cite{ChRaRuSm89,Te91}, we have
\begin{equation}\label{GMFPT03}
F_{ij}(t)+F_{ji}(t)=2\,E_t\,R_{ij}(t)\,.
\end{equation}
Therefore, Eq.~(\ref{GMFPT02}) can be rewritten as
\begin{equation}\label{GMFPT04}
T_{\rm tot}(t)=E_{t}\,\sum_{j= 1}^{N_t}\sum_{\stackrel{i=1}{ i \ne
j}}^{N_t}R_{ij}(t)\,.
\end{equation}
Thus, if one knows how to determine the effective resistance, then
we have a method to find $\langle F \rangle_t$. Then, the question
of determining $\langle F \rangle_t$ is reduced to computing the
total resistance $R_{\rm tot}(t)$ between all pairs of nodes in the
resistor networks:
\begin{equation}\label{GMFPT05}
R_{\rm tot}(t)=\sum_{j= 1}^{N_t}\sum_{\stackrel{i=1}{ i \ne
j}}^{N_t}R_{ij}(t)\,.
\end{equation}

According to the structure of the Koch networks, it is obvious that
the effective resistance between any two nodes is exactly
$\frac{2}{3}$ times the usual shortest-path distance between the
corresponding nodes, i.e.,
\begin{equation}\label{GMFPT06}
R_{ij}(t)=\frac{2}{3}d_{ij}(t)\,,
\end{equation}
where $d_{ij}(t)$ is the shortest distance between nodes $i$ and $j$
in $K_{m,t}$. Equation~(\ref{GMFPT06}) can be interpreted as
follows. By construction, the Koch networks consist of triangles;
moreover, no edge lies in more than one triangle. Then, for any
couple of nodes $i$ and $j$ in $K_{m,t}$, the shortest path between
them is unique. It is easy to see that the effective resistance
between two nodes directly connected by an edge in the shortest path
of $i$ and $j$ is $\frac{2}{3}$, which is in fact equal to the
effective resistance between two nodes of a triangle. And the
$R_{ij}(t)$ can be regarded as the sum of effective resistance of
$d_{ij}(t)$ conductors in series, each of which has a effective
resistance of $\frac{2}{3}$.

Then, to obtain $\langle F \rangle_t$, we need only to calculate the
total of shortest distances between all node pairs, denoted by
$D_{\rm tot}(t)$, namely
\begin{equation}\label{GMFPT07}
D_{\rm tot}(t)=\sum_{i\neq
 j}\sum_{j=1}^{N_t}d_{ij}(t)\,.
\end{equation}
It is then obvious to have
\begin{equation}\label{GMFPT08}
R_{\rm tot}(t)=\frac{2}{3}\,D_{\rm tot}(t)\,.
\end{equation}
Hence, all that is left to find $\langle F \rangle_t$ is to evaluate
$D_{\rm tot}(t)$.
% notice that since for an arbitrary pair of nodes $i$ and $j$ ($i
%\neq j$) we have $d_{ij}(t)=d_{ji}(t)$, in Eq.~(\ref{GMFPT07}) we
%only count $d_{ij}(t)$ or $d_{ji}(t)$, not both.

According to our previous result~\cite{ZhGaChZhZhGu10}, we can
easily obtain the closed-form expression for $D_{\rm tot}(t)$:
\begin{eqnarray}\label{GMFPT09}
&\quad& D_{\rm tot}(t)\nonumber \\
&=& \frac{2(3m+1)^{t-1}}{3} \left
[3m+5+(24mt+24m+4)(3m+1)^{t}\right].\nonumber \\
\end{eqnarray}
Combining above-obtained results, we arrive at the explicit solution
to $\langle F \rangle_t$:
\begin{eqnarray}\label{GMFPT10}
&\quad& \langle F \rangle_t\nonumber \\
&=&\frac{2}{3}\frac{1}{N_t(N_t-1)}E_t\,D_{\rm tot}(t)\nonumber \\
&=&\frac{2(3m+1)^{t-1}}{6(3m+1)^{t}+3}\left
[3m+5+(24mt+24m+4)(3m+1)^{t}\right],\nonumber \\
\end{eqnarray}
which can be expressed in terms of network order $N_t$ as
\begin{eqnarray}\label{GMFPT11}
\langle F \rangle_t &=&
\frac{1}{3(3m+1)}\frac{N_t-1}{N_t}\Big[3m+5+(N_t-1)\nonumber
\\&\quad& \Big(\frac{12m\ln(N_t-1)-12m\ln2}{\ln(3m+1)}+12m+2\Big)\Big] \,.\nonumber \\
\end{eqnarray}

Equation~(\ref{GMFPT11}) uncovers the exact dependence relation of
GMFPT on network order $N_t$ and parameter $m$. For large systems,
i.e., $N_t\rightarrow \infty$, we have following expression for the
leading term of $\langle T \rangle_t$:
\begin{equation}\label{GMFPT12}
\langle F \rangle_t \sim
\frac{4m}{(3m+1)\ln(3m+1)}(N_t-1)\ln(N_t-1)\,,
\end{equation}
which is in consistent with the general result given
in~\cite{TeBeVo09}. Thus, similar to the behavior of AST obtained in
the previous section, in the large limit of $t$, the GMFPT grows
with network order $N_t$ as $N_t \ln N_t$, which is independent of
$m$ and thus shows that the structure heterogeneity of the networks
has no substantial impact on the scaling of GMFPT. The sameness for
the leading behavior between $\langle T \rangle_t$ and $\langle F
\rangle_t$ implies that the $N_t \ln N_t$ scaling of $\langle T
\rangle_t$ from a hub node to all other nodes is a representative
feature for information sending in the Koch networks.

The $N_t \ln N_t$ behavior found for both the AST and GMFPT can be
understood from the following heuristic explanations. The couples of
nodes farthest apart (between each other and from the hub due to its
centrality) provide the leading contribute for the related
MFPTs~\cite{ZhLiMa10}. On the other hand, for the ART related to the
trapping problem with the trap fixed on a hub node, since the hub is
relatively easy to reach for most nodes, the ART is relatively small
and contributes little to GMFPT, see also~\cite{AgBuMa10}.

Note that if the information sender is positioned at an arbitrary
non-hub node in networks $K_{m,t}$. The AST from the sender to all
other nodes also follows the scaling $N_t \ln N_t$. Because in most
of this case, the information must be first delivered to a hub node
in a time at most proportional to network order
$N_t$~\cite{ZhZhXiChLiGu09}, then the piece of information proceeds
to be sent, until it reaches the receiver after an average transmit
time $N_t \ln N_t$ as shown in the previous section. To confirm
this, we have computed analytically the AST for the sender located
at new neighbor of hub node $1$ created at step $t$, and obtained
the same expression as Eqs.~(\ref{TMFPT22}) and (\ref{GMFPT12}).

\section{Conclusions}

We have studied random walks on the Koch network family, exhibiting
synchronously scale-free and small-world behaviors. We first
concentrated on a specific case for random walks from a hub node to
all other nodes, and obtained explicitly the formula for AST from
this most connected node to different target nodes, which varies
with network order $N$ as $N \ln N$, larger than the ART from all
other nodes to the hub. Then we continued to derive the GMFPT
between two arbitrary nodes averaged over all node couples in the
Koch networks, which can be regarded as the average of MFPTs from a
uniformly-selected starting point to all other nodes in the
networks. We presented that in the limit of large network order $N$,
the GMFPT also scales approximatively with $N$ as $N \ln N$. This
identity of scalings between the AST and GMFPT indicates that the
ability (efficiency) of hub nodes sending information is the same as
that the average efficiency of all nodes in the Koch networks,
showing that the sending efficiency measured by AST is not
sensitively influenced by the position of information sender and the
structural heterogeneity of the networks. Finally, it should be
mentioned that we only studied a particular family of scale-free
networks, whether the conclusion also holds for other scale-free
networks, even general networks, needs further investigation in the
future.

\section*{Acknowledgment}

This research was supported by the National Natural Science
Foundation of China under Grants No. 61074119 and the
Shanghai Leading Academic Discipline Project No. B114. %S. Y. G. also acknowledges the support by Fudan's Undergraduate Research Opportunities Program.

%%%%%%%%%%%%%%%%%%%%%%%%%%%%%%%%%%%%%%%%%%%%%%%%%%%%%%%%%%%%%%%%%
%%%%%%%%%%%%%%%%%%%%%%%%%%%%%%%%%%%%%%%%%%%%%%%%%%%%%%%%%%%%%%%%%


\begin{thebibliography}{10}
%% 4


\bibitem{AlBa02} R. Albert and A.-L. Barab\'asi,
      %Statistical mechanics of complex networks,
       Rev. Mod. Phys. {\bf 74}, 47 (2002).

\bibitem{DoMe02} S. N. Dorogvtsev and J.F.F. Mendes,
%Evolution of networks,
Adv. Phys. {\bf 51}, 1079 (2002).


\bibitem{Ne03} M. E. J. Newman,
%The structure and function of complex networks,
SIAM Rev. {\bf 45}, 167 (2003).

\bibitem{BoLaMoChHw06}
S. Boccaletti, V. Latora, Y. Moreno, M. Chavezf, and D.-U. Hwanga,
 %Complex networks: Structure and dynamics.
Phy. Rep. {\bf 424}, 175 (2006).

\bibitem{WaSt98}
D.J. Watts and H. Strogatz,
       %Collective dynamics of `small-world' networks,
        Nature (London) {\bf 393}, 440 (1998).

\bibitem{BaAl99}
A.-L. Barab\'asi and R. Albert,
      %Emergence of scaling in random networks,
       Science {\bf 286}, 509 (1999).

\bibitem{DoGoMe08}
S. N. Dorogovtsev, A. V. Goltsev and J. F. F. Mendes,
      %Critical phenomena in complex networks,
       Rev. Mod. Phys. {\bf 80}, 1275 (2008).

\bibitem{PaVe01}
R. Pastor-Satorras and A. Vespignani, Phys. Rev. Lett. {\bf 86},
3200 (2001).

\bibitem{CaNeStWa00}
D. S. Callaway, M. E. J. Newman,  S. H. Strogatz, and D. J. Watts,
%   Network Robustness and Fragility: Percolation on Random Graphs
Phys. Rev. Lett. {\bf 85}, 5468 (2000).

\bibitem{SaPa05}
F. C. Santos and J. M. Pacheco,
%Scale-Free Networks Provide a Unifying Framework for the Emergence of Cooperation
Phys. Rev. Lett. {\bf 95}, 098104 (2005).

\bibitem{SaSaPa08}
F. C. Santos, M. D. Santos, and J. M. Pacheco,
% Social diversity promotes the emergence of cooperation in public goods games
Nature (London) {\bf 454}, 213 (2008).

\bibitem{ArDiKuMoZh08}
A. Arenas, A. D\'iaz-Guilera, J. Kurths, Y. Moreno, and C. S. Zhou,
Phy. Rep. {\bf 469}, 93 (2008).

\bibitem{KiCaHaAr08}
A. Kittas, S. Carmi, S. Havlin, and P. Argyrakis, EPL {\bf 84},
40008 (2008).

\bibitem{ZhQiZhXiGu09}
Z. Z. Zhang, Y. Qi, S. G. Zhou, W. L. Xie, and J. H. Guan, Phys.
Rev. E {\bf 79}, 021127 (2009).

\bibitem{ZhGuXiQiZh09}
Z. Z. Zhang, J. H. Guan, W. L. Xie, Y. Qi, and S. G. Zhou, EPL, {\bf
86}, 10006 (2009).

\bibitem{AgBu09}
E. Agliari and R. Burioni, Phys. Rev. E {\bf 80}, 031125 (2009).

\bibitem{TeBeVo09}
V. Tejedor, O. B\'enichou, and R. Voituriez, Phys. Rev. E {\bf 80},
065104(R) (2009).

\bibitem{AgBuMa10}
E. Agliari, R. Burioni, and A. Manzotti, Phys. Rev. E {\bf 82},
011118 (2010).

\bibitem{Re01}
S. Redner, \emph{A Guide to First-Passage Processes} (Cambridge
University Press, Cambridge, 2001).

\bibitem{NoRi04}
J. D. Noh and H. Rieger, Phys. Rev. Lett. {\bf 92}, 118701 (2004).

\bibitem{ChBa07}
C. Chennubhotla and I. Bahar, PLoS Comput. Biol. {\bf 3}, 1716
(2007).

\bibitem{Bobe05}
E. Bollt and  D. ben-Avraham, New J. Phys. {\bf 7}, 26 (2005).

\bibitem{ZhZhXiChLiGu09}
Z. Z. Zhang,  S. G. Zhou, W. L. Xie, L. C. Chen, Y. Lin,  and J. H.
Guan, Phys. Rev. E {\bf 79}, 061113 (2009).

\bibitem{ZhGaChZhZhGu10}
Z. Z. Zhang, S. Y. Gao, L. C. Chen, S. G. Zhou, H. J. Zhang, and J.
H. Guan, J. Phys. A {\bf 43}, 395101 (2010).

\bibitem{MeKl04}
R. Metzler and J. Klafter, J. Phys. A {\bf 37}, R161 (2004).

\bibitem{NoRi04a}
J. D. Noh and H. Rieger, Phys. Rev. E {\bf 69}, 036111 (2004).

\bibitem{SoRebe05}
V. Sood, S. Redner, and D. ben-Avraham, J. Phys. A {\bf 38}, 109
(2005).

\bibitem{BaLo06}
A. Baronchelli and V. Loreto,
%Ring structures and mean first passage time in networks
Phys. Rev. E {\bf 73}, 026103 (2006).

\bibitem{CoBeTeVoKl07}
S. Condamin, O. B\'enichou, V. Tejedor, R. Voituriez, and J.
Klafter, Nature (London) {\bf 450}, 77 (2007).

\bibitem{GaSoHaMa07}
L. K. Gallos, C. Song, S. Havlin, and H. A. Makse, Proc. Natl. Acad.
Sci. U.S.A. {\bf 104}, 7746 (2007).

\bibitem{CoTeVoBeKl08}
S. Condamin, V. Tejedor, R. Voituriez, O. B\'enichou and J. Klafter,
Proc. Natl. Acad. Sci. U.S.A. {\bf 105}, 5675 (2008).

\bibitem{BeChKlMeVo10}
O. B\'enichou, C. Chevalier, J. Klafter, B. Mayer, and R. Voituriez,
Nat. Chem. {\bf 2}, 472 (2010).

\bibitem{BeGrLeLoVo10}
O. B\'enichou, D. Grebenkov, P. Levitz, C. Loverdo, and R.
Voituriez, Phys. Rev. Lett. {\bf 105}, 150606 (2010).


\bibitem{LaVaMeVa87}
A. Lakhtakia, V. K. Varadan, R. Messier, and V. V. Varadan, J. Phys.
A {\bf 20}, 3537 (1987). %: Math. Gen.


\bibitem{HaBe87}
S. Havlin and D. ben-Avraham, Adv. Phys. {\bf 36}, 695 (1987).


\bibitem{ZhZhZhYiGu09}
Z. Z. Zhang, Y. C. Zhang, S. G. Zhou, M. Yin, and J. H. Guan, J.
Math. Phys. {\bf 50}, 033514 (2009).

\bibitem{AlFi99}
D. Aldous and J. Fill, Reversible Markov chains and random walks on
graphs, 1999, http://www.stat.berkeley.edu/~aldous/RWG/Chap2.pdf

\bibitem{SaDoMe08}
A. N. Samukhin, S. N. Dorogovtsev, and J. F. F. Mendes, Phys. Rev. E
{\bf 77}, 036115 (2008).


\bibitem{ChRaRuSm89}
A. K. Chandra, P. Raghavan, W. L. Ruzzo, and R. Smolensky, in
\emph{Proceedings of the 21st Annnual ACM Symposium on the Theory of
Computing} (ACM Press, New York, 1989), pp. 574-586.

\bibitem{Te91}
P. Tetali, J. Theor. Probab. {\bf 4}, 101 (1991).

\bibitem{DoSn84}
P. G. Doyle and J. L. Snell, \emph{Random Walks and Electric
Networks} (The Mathematical Association of America, Oberlin, OH,
1984); e-print arXiv:math.PR/0001057.

\bibitem{ZhLiMa10}
Z. Z. Zhang, Y. Lin, and Y. J. Ma (unpublished).%J. Phys. A {\bf 79}, 021127 (2009).




\end{thebibliography}
\end{document}